\documentclass[aps,prl,amsmath,amssymb,reprint]{revtex4-2}
\usepackage[utf8]{inputenc}
\usepackage[T1]{fontenc}
\usepackage{graphicx}
\usepackage{color}
\usepackage{url}
\usepackage[hidelinks]{hyperref}

\renewcommand*{\vec}[1]{\mathbf{#1}}

\graphicspath{{./figures/}}

\begin{document}

\title{Determining the chemical potential via universal density functional learning}

\author{Florian Sammüller}
\email{florian.sammueller@uni-bayreuth.de}
\author{Matthias Schmidt}
\email{matthias.schmidt@uni-bayreuth.de}
\affiliation{Theoretische Physik II, Physikalisches Institut, Universität Bayreuth, D-95447 Bayreuth, Germany}

\date{\today}

\begin{abstract}
  We demonstrate that the machine learning of density functionals allows one to determine simultaneously the equilibrium chemical potential across simulation datasets of inhomogeneous classical fluids.
  Minimization of a loss function based on an Euler-Lagrange equation yields both the universal one-body direct correlation functional, which is represented locally by a neural network, as well as the system-specific unknown chemical potential values.
  The method can serve as an efficient alternative to conventional computational techniques of measuring the chemical potential.
  It also facilitates using canonical data from Brownian dynamics, molecular dynamics, or Monte Carlo simulations as a basis for constructing neural density functionals, which are fit for accurate multiscale prediction of soft matter systems in equilibrium.
\end{abstract}

\maketitle

Various different approaches to understanding the chemical potential $\mu$ have been put forward \cite{CookUnderstandingChemicalPotential1995,BaierleinElusiveChemicalPotential2001,JobChemicalPotentialQuantity2006}.
From a fundamental statistical mechanical perspective, $\mu$ acts as an additive constant to the potential energy of each particle \cite{HansenTheorySimpleLiquids2013}.
The corresponding statistical weight for each microstate of $N$ particles is $\mathrm{e}^{-\beta(H-\mu N)}/\Xi$, where $H$ denotes the Hamiltonian, $\Xi$ is the grand partition sum, and $\beta = 1 / (k_B T)$ with absolute temperature $T$ and the Boltzmann constant $k_B$.
In computer simulations, Widom's test particle insertion method \cite{WidomTopicsTheoryFluids1963,JacksonPotentialDistributionMethod1964} and its powerful variants and generalizations, see e.g.\ Refs.~\cite{NezbedaNewVersionInsertion1991,PowlesNondestructiveMoleculardynamicsSimulation1982,ShingChemicalPotentialComputer1981,ShingChemicalPotentialNonideal1983,SmitCalculationChemicalPotential1989,KofkeQuantitativeComparisonOptimization1997,LuAppropriateMethodsCombine2003}, allow one to express the chemical potential as a canonical average, which circumvents the need to represent the grand ensemble directly \cite{FrenkelUnderstandingMolecularSimulation2023}.
Specifically, the nontrivial excess (over ideal gas) chemical potential is obtained as $\beta \mu_\mathrm{exc} = -\ln \langle \exp(-\beta \epsilon) \rangle$, where the angle brackets denote a canonical thermal equilibrium average with $N$ particles, and $\epsilon$ is the energy change caused by adding an $(N+1)$th particle.
Generating values for $\epsilon$ is readily performed in simulations via virtual particle insertions.

Testing and developing methods that are applicable to hard sphere systems \cite{SantosStructuralThermodynamicProperties2020,RoyallColloidalHardSpheres2024} is important due to the particular features arising from purely entropic behavior \cite{AttardSimulationChemicalPotential1993,HeyesChemicalPotentialTest2016,HeyesChemicalPotentialTest2018,DavidchackChemicalPotentialSurface2022}.
Via analyzing colloidal particle coordinates obtained from optical microscopy, \citeauthor{DullensDirectMeasurementFree2006} \cite{DullensDirectMeasurementFree2006} determined the chemical potential in their experimental hard sphere system.
Computational particle insertion can be performed in a variety of ways, including via growth from an added fourth dimension \cite{BelloniNonequilibriumHybridInsertion2019}.
Widom insertion also generalizes to more complex systems, such as polymers \cite{WildingAccurateMeasurementsChemical1994}.
Furthermore, the chemical potential was shown to be accessible via adaptive resolution \cite{WangGrandCanonicallikeMolecularDynamicsSimulations2013,AgarwalChemicalPotentialLiquids2014,DelleSiteMolecularDynamicsOpen2019,GholamiThermodynamicRelationsCoupling2021} and parallel computing techniques \cite{DalyMassivelyParallelChemical2012,ZhaoGPUspecificAlgorithmsImproved2023}.
Its role in relevant questions of phase coexistence \cite{BinderVanWaalsLoop2012} and of finite size effects \cite{SiepmannFinitesizeCorrectionsChemical1992} was addressed.
While much of the above work concerns bulk fluids, several insightful studies were specifically targeted at spatially inhomogeneous systems \cite{HeinbuchApplicationWidomsTest1987,PeregoChemicalPotentialCalculations2016,PeregoChemicalPotentialCalculations2018,HeidariSpatiallyResolvedThermodynamic2018,UstinovEfficientChemicalPotential2017}, which are pertinent to tackle the wide range of adsorption and interfacial phenomena in soft matter.

Machine learning is currently applied across a rapidly growing scope of problems in statistical mechanics, ranging from the characterization of soft matter \cite{CleggCharacterisingSoftMatter2021}, reverse-engineering of self-assembly \cite{DijkstraPredictiveModellingMachine2021}, and local structure detection \cite{BoattiniUnsupervisedLearningLocal2019} to the investigation of many-body potentials \cite{Campos-VillalobosMachineLearningManybody2021,Campos-VillalobosMachinelearningEffectiveManybody2022}.
Notably, a surge of machine learning techniques has recently been developed in the context of classical density functional theory \cite{Santos-SilvaNeuralnetworkApproachModeling2014,LinClassicalDensityFunctional2019,LinAnalyticalClassicalDensity2020,CatsMachinelearningFreeenergyFunctionals2021,QiaoEnhancingGasSolubility2020,YatsyshinPhysicsconstrainedBayesianInference2022,Malpica-MoralesPhysicsinformedBayesianInference2023,FangReliableEmulationComplex2022,SimonMachineLearningDensity2024,SimonOrientationalStructureModel2025,DeLasHerasPerspectiveHowOvercome2023,ZimmermannNeuralForceFunctional2024,SammullerNeuralFunctionalTheory2023,SammullerNeuralFunctionalTheory2023a,SammullerWhyNeuralFunctionals2024,SammullerNeuralDensityFunctionals2024,SammullerHyperdensityFunctionalTheory2024,SammullerWhyHyperdensityFunctionals2025,KampaMetadensityFunctionalTheory2025,SammullerNeuralDensityFunctional2025,BuchananMachineLearningPredicts2025,YangHighDimensionalOperatorLearning2024,BuiLearningClassicalDensity2025,BuiFirstPrinciplesApproach2025,BuiDielectrocapillarityExquisiteControl2025,GlitschNeuralDensityFunctional2025,DijkmanLearningNeuralFreeEnergy2025,KelleyBridgingElectronicClassical2024,PanNeuralOperatorsForward2025}.
Besides the prime goal of approximating the underlying density functional mappings, various methods have also proven feasible for addressing complex observables \cite{SammullerHyperdensityFunctionalTheory2024,SammullerWhyHyperdensityFunctionals2025}, inverse design problems \cite{DeLasHerasPerspectiveHowOvercome2023,ZimmermannNeuralForceFunctional2024,KampaMetadensityFunctionalTheory2025} and the dynamics in nonequilibrium systems \cite{DeLasHerasPerspectiveHowOvercome2023,ZimmermannNeuralForceFunctional2024}.

We recall that classical density functional theory is founded on a formally exact variational principle of the grand potential \cite{EvansNatureLiquidvapourInterface1979}, which is expressed as a functional of the density profile $\rho(\vec{r})$, where $\vec{r}$ denotes (generic) position.
Analogously to its quantum mechanical counterpart \cite{HohenbergInhomogeneousElectronGas1964,PedersonMachineLearningDensity2022}, this theoretical underpinning facilitates the development of highly accurate and efficient machine learning techniques.
Specifically, the rigorous functional relationships lend themselves very naturally to be represented by neural networks, and they enable a tight integration of modern computational tools, in particular automatic differentiation, for analyzing a wide range of physical behavior \cite{StierleClassicalDensityFunctional2024}.

\begin{figure}[t]
  \includegraphics{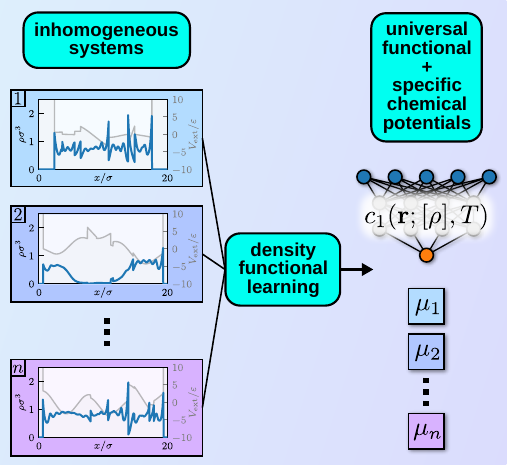}
  \caption{
    Workflow for determining chemical potential values across a dataset of $n$ inhomogeneous systems via density functional learning.
    Data from one simulation run $k$ consists of the temperature value $T_k$, the external potential $V_\mathrm{ext}^{(k)}(\vec{r})$ (gray lines), and the density profile $\rho_k(\vec{r})$ (blue lines); in one spatial dimension or for planar symmetry $\vec{r} = x$.
    Supervised training based on the loss function \eqref{eq:EL_loss} determines both the universal one-body direct correlation functional $c_1(\vec{r}; [\rho], T)$, which is represented locally as a neural network (see Appendix~A), as well as the system-specific chemical potential values $\mu_k$.
  }
  \label{fig:overview}
\end{figure}

All currently available methods for learning classical density functionals rely on the explicit access to grand canonical simulation data, which entails known (prescribed) chemical potential values for each system in a given dataset of inhomogeneous fluid equilibria.
In this Letter, we present an augmentation of density functional learning, which does not require the chemical potential values to be known.
We show that hitherto unknown chemical potentials can instead be determined simultaneously for all systems during the training of neural density functionals without any additional computational effort and crucially without resorting back to many-body simulation techniques.
The benefit of our approach is twofold:
i) The method constitutes a high-throughput scheme for measuring chemical potentials across large datasets of inhomogeneous fluids.
ii) Constructing neural density functionals is no longer restricted to using simulation data with full grand canonical information.
Both of these aspects are relevant in practice, and we describe possible applications below.
Making density functional learning more versatile is particularly pertinent in light of the arguably much more widespread use of simulation techniques in the $NVT$ and $NpT$ ensembles compared to grand canonical schemes.
Concretely, our method facilitates the direct utilization of data from molecular or Brownian dynamics simulations for training neural functionals, provided that ensemble differences in the acquired density profiles remain negligible.
An illustration of the central workflow is shown in Fig.~\ref{fig:overview}.

We first give an overview of fundamentals.
A central object of classical density functional theory is the one-body direct correlation functional $c_1(\vec{r}; [\rho], T)$.
Besides its functional dependence on the density profile $\rho(\vec{r})$, denoted by brackets, the temperature $T$ arises as an additional parametric dependence for general (beyond hard-core) fluid types.
We keep the following presentation general, but drop the dependence on $T$ if athermal systems or applications for fixed temperature are considered.
For a given type of fluid, as specified by its intrinsic Hamiltonian, $c_1(\vec{r}; [\rho], T)$ is a unique functional, independent of the external potential \cite{EvansNatureLiquidvapourInterface1979}.
It is hence universally applicable for predicting equilibrium states of the considered fluid in arbitrary inhomogeneous environments \footnote{The existence of universal density functionals has originally been demonstrated in seminal work for groundstates of quantum many-body systems \cite{HohenbergInhomogeneousElectronGas1964}. This formal structure holds analogously for the case of classical fluids~\cite{EvansNatureLiquidvapourInterface1979}.}.
Assuming that the universal functional $c_1(\vec{r}; [\rho], T)$ is known, self-consistently solving a corresponding Euler-Lagrange equation,
\begin{equation}
  \label{eq:EL}
  \rho(\vec{r}) = \exp\left( -\beta V_\mathrm{ext}(\vec{r}) + \beta \mu + c_1(\vec{r}; [\rho], T) \right),
\end{equation}
yields a prediction for the density profile $\rho(\vec{r})$ that is generated by the external potential $V_\mathrm{ext}(\vec{r})$ at the given statepoint.
Obtaining a numerical solution of Eq.~\eqref{eq:EL} is typically both straighforward and efficient, even for very large systems with high spatial resolution \cite{SammullerNeuralFunctionalTheory2023}.
Due to the fundamental role of $c_1(\vec{r}; [\rho], T)$, recent machine learning approaches have focused on representing this functional in terms of a neural network; an overview of feasible techniques for the construction of neural density functionals is given in Appendix~A.

Equation~\eqref{eq:EL} is also key for the analysis of a dataset of inhomogeneous fluid states, and it will hence serve as the basis for training neural density functionals.
As a template, we first consider grand canonical simulations, where the thermodynamic statepoint is characterized by known system-specific values of $T$ and $\mu$.
Upon imposing a spatially inhomogeneous external potential $V_\mathrm{ext}(\vec{r})$, the resulting density profile $\rho(\vec{r})$ becomes inhomogeneous in general.
Its microscopic definition is $\rho(\vec{r}) = \left\langle \sum_i \delta(\vec{r} - \vec{r}_i) \right\rangle$, where the angle brackets indicate an equilibrium grand ensemble average, the sum runs over all particles $i = 1, \dots, N$ with positions $\vec{r}_i$, and $\delta(\cdot)$ denotes the Dirac distribution.
This average can be measured in simulations via standard histogram-counting methods or with more advanced reduced-variance (force-sampling) algorithms \cite{BorgisComputationPairDistribution2013,DeLasHerasBetterCountingDensity2018,RotenbergUseForceReduced2020}.

We henceforth consider $n$ individual simulation runs which contribute to the dataset and assign to each result the label $k = 1, \dots, n$ to keep track of the respective simulation origin.
Rearranging Eq.~\eqref{eq:EL} yields
\begin{equation}
  \label{eq:c1_EL}
  c_1^{(k)}(\vec{r}) = \ln \rho_k(\vec{r}) + \beta_k V_\mathrm{ext}^{(k)}(\vec{r}) - \beta_k \mu_k,
\end{equation}
which makes all quantities on the right-hand side known.
Via Eq.~\eqref{eq:c1_EL}, the one-body direct correlation function $c_1^{(k)}(\vec{r})$ that corresponds to the grand canonical simulation run $k$ is readily computed in a post-processing step on the acquired simulation data for each position $\vec{r}$ where $\rho_k(\vec{r}) > 0$.
The relationship $\rho_k(\vec{r}), T_k \rightarrow c_1^{(k)}(\vec{r})$ can hence be constructed explicitly for all simulated systems if all values of $\mu_k$ are known.
As classical density functional theory ascertains the universality of this functional mapping, training a neural network on the thus acquired dataset is feasible in order to obtain a highly accurate numerical representation of $c_1(\vec{r}; [\rho], T)$.
Specifically, we use the local one-body correlation learning scheme \cite{SammullerNeuralFunctionalTheory2023,SammullerNeuralFunctionalTheory2023a,SammullerWhyNeuralFunctionals2024,SammullerNeuralDensityFunctional2025} and provide further details in Appendix~A.

We now modify the density functional learning in order to not require the chemical potential values to be known.
Hence, our aim is to work solely with input data $T_k$, $V_\mathrm{ext}^{(k)}(\vec{r})$, and $\rho_k(\vec{r})$, as are directly available already in canonical simulations.
Our treatment of the set of the then unknown $\mu_k$ is similar to treating the parameters of the neural network: both are to be determined during the training procedure.
Additionally to the neural network which represents $c_1(\vec{r}; [\rho], T)$, we therefore prepare an array of $n$ numerical variables which correspond individually to the chemical potentials of the $n$ simulations.
The neural network parameters and the chemical potential values are simultaneously optimized via standard backpropagation, as illustrated in Fig.~\ref{fig:overview} and as described in the following.

The method rests on the fact that $\beta \mu$ merely constitutes an additive offset to $c_1(\vec{r}; [\rho], T)$ in Eq.~\eqref{eq:EL}, which is identical for all positions $\vec{r}$ in a given system.
Thus, Eq.~\eqref{eq:EL} serves as a simultaneous matching condition for both $c_1(\vec{r}; [\rho_k], T_k)$ and $\mu_k$ for each simulation run $k$ in the dataset.
We therefore define the loss function
\begin{equation}
  \label{eq:EL_loss}
  \begin{split}
    &\mkern-8mu \mathcal{L}_\mathrm{EL} =\\
    &\mkern-8mu \sum_{k=1}^n \left\Vert \ln \rho_k(\vec{r}) + \beta_k V_\mathrm{ext}^{(k)}(\vec{r}) - \beta_k \mu_k - c_1(\vec{r}; [\rho_k], T_k) \right\Vert_2^2,
  \end{split}
\end{equation}
which measures via the spatial norm $\Vert \cdot \Vert_2$ how strongly the Euler-Lagrange equation~\eqref{eq:EL} is violated across all members $k = 1, \dots, n$ of the dataset.
Both the functional form of $c_1(\vec{r}; [\rho], T)$ as well as the values $\mu_k$ thereby enter as predictions and hence as the targets to be optimized.
In practice, all profiles are provided numerically at discretized positions~$\vec{r}$.
The implementation of Eq.~\eqref{eq:EL_loss} is straightforward \cite{SammullerNeuralMu2025} as its form allows utilizing the standard mean squared error loss function readily available in common machine learning libraries.

Clearly $\mathcal{L}_\mathrm{EL} \geq 0$, with equality holding only if Eq.~\eqref{eq:EL} is satisfied across the entire dataset.
Hence, minimization of the loss function \eqref{eq:EL_loss} determines simultaneously the values $\mu_k$ and the functional $c_1(\vec{r}; [\rho], T)$, where the latter is represented by the neural network parameters.
We emphasize two central points which rationalize the successful convergence (see below) of this minimization:
i) The functional mapping $c_1(\vec{r}; [\rho], T)$ is universal for the considered fluid type \cite{EvansNatureLiquidvapourInterface1979}, i.e.\ it is independent of chemical and external potential.
ii) For each simulation run $k$, while the functional relationship contributes individually at every (discretized) position $\vec{r}$ throughout the system to the loss function \eqref{eq:EL_loss}, the chemical potential necessarily corresponds to the same system-specific value $\mu_k$.
Thus, the values $\mu_k$ can be recovered consistently across the whole dataset.

As a practical detail, the minimization of Eq.~\eqref{eq:EL_loss} determines the resulting functional $c_1(\vec{r}; [\rho], T)$ and the set of values $\beta_k \mu_k$ only up to a global additive constant $\tilde{c}_1$.
This follows trivially from substituting $\beta_k \mu_k \rightarrow \beta_k \mu_k + \tilde{c}_1$ and $c_1(\vec{r}; [\rho_k], T_k) \rightarrow c_1(\vec{r}; [\rho_k], T_k) - \tilde{c}_1$ in Eq.~\eqref{eq:EL_loss}, which leaves $\mathcal{L}_\mathrm{EL}$ unchanged.
This global offset can be corrected a posteriori by demanding that $c_1(\vec{r}; [\rho = 0], T) = 0$.
In applications, the neural functional must hence be evaluated for vanishing density input and this offset needs to be subtracted from the inferred value at target density $\rho(\vec{r})$ to obtain the correct result for $c_1(\vec{r}; [\rho], T)$.
Similarly, the recovered values $\mu_k$ have to be corrected by the same (negative) offset.

\begin{figure}[t]
  \includegraphics{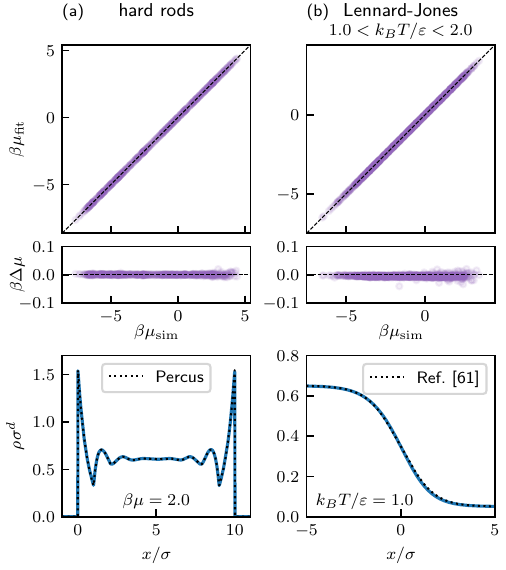}
  \caption{
    Predictions for the scaled chemical potential values $\beta \mu_\mathrm{fit}$ (top row) for datasets of hard rods in $d = 1$ dimension (a) and the truncated Lennard-Jones fluid in $d = 3$ dimensions and planar symmetry (b).
    The Lennard-Jones fluid is addressed via thermal training \cite{SammullerNeuralDensityFunctional2025} in the temperature range $1.0 < k_B T / \varepsilon < 2.0$.
    The results show excellent agreement with the simulation reference $\beta \mu_\mathrm{sim}$, see the consistently small difference $\beta \Delta \mu = \beta \mu_\mathrm{fit} - \beta \mu_\mathrm{sim}$ (middle row).
    The trained neural correlation functionals $c_1(x; [\rho])$ (a) and $c_1(x; [\rho], T)$ (b) are directly applicable for the calculation of density profiles (bottom row, solid blue lines).
    We consider hard-wall confinement in a planar slit of length $10 \sigma$ for hard rods (a) and compare the prediction to Percus' exact density functional result \cite{PercusEquilibriumStateClassical1976} (dotted black line).
    For the thermally trained functional of the Lennard-Jones fluid (b), we predict phase coexistence and show an exemplary density profile of the liquid-gas interface, which is compared to the neural result of Ref.~\cite{SammullerNeuralDensityFunctional2025} (dotted black line).
  }
  \label{fig:1D-hard-rods_3D-LJ-Tvar}
\end{figure}

For an initial proof of principle, we turn to (athermal) one-dimensional hard rods with particle size~$\sigma$, where neural density functionals deliver highly accurate results \cite{SammullerNeuralFunctionalTheory2023a,SammullerWhyNeuralFunctionals2024}, and with Percus' exact analytical free energy functional \cite{PercusEquilibriumStateClassical1976} serving as the ultimate benchmark.
While we use grand canonical Monte Carlo simulations, which comprise the full dataset $\mu_k$, $T_k$, $\rho_k(x)$, and $V_\mathrm{ext}^{(k)}(x)$, where $k = 1, \dots, 512$ and $x$ denotes one-dimensional position, we deliberately disregard all values $\mu_k$ during training.
We proceed instead via the above methodology using the Euler-Lagrange-based loss function \eqref{eq:EL_loss} to determine these values along with the functional $c_1(x; [\rho])$.
Our goal is hence to recover via the training all values $\mu_k$, which can be verified a posteriori against the (known) grand canonical Monte Carlo input parameters.
We observe that the values $\mu_k$ converge already after few training epochs.
Comparison with the available reference data for $\mu_k$, upon correcting the global offset $\tilde{c}_1$, confirms excellent agreement, as shown in Fig.~\ref{fig:1D-hard-rods_3D-LJ-Tvar}(a), which corroborates that unknown chemical potentials can be recovered via the present procedure.
Moreover, utilizing the resulting neural functional $c_1(x; [\rho])$ in predictions yields density profiles which closely match those obtained from the exact Percus functional, as verified for the case of hard-wall confinement.

In order to demonstrate the prowess of the method, we next consider the three-dimensional truncated Lennard-Jones fluid, which is specified by the pairwise interparticle potential $\phi(r) = 4 \varepsilon [(\sigma / r)^{12} - (\sigma / r )^6]$ for particle distance $r < r_c = 2.5\sigma$ and $\phi(r) = 0$ otherwise, where $\varepsilon$ and $\sigma$ set the energy and length scale.
The grand canonical reference dataset consists of $n = 882$ simulation results at varying temperatures $1.0 < k_B T / \varepsilon < 2.0$, such that the neural functional $c_1(x; [\rho], T)$ acquires additional parametric dependence on temperature \cite{SammullerNeuralDensityFunctional2025}.
Inhomogeneous external potentials are imposed in planar geometry with $x$ denoting the spatial coordinate along the inhomogeneous direction.
As before, all values $\mu_k$ are, though available, crucially not utilized in the training.
Instead, the chemical potentials are recovered during density functional learning, and the resulting values closely match the reference also in this case, as shown in Fig.~\ref{fig:1D-hard-rods_3D-LJ-Tvar}(b).
Additionally, the thermally trained neural functional is fit for application, which we demonstrate for the prototypical problem of determining the interfacial density profile of coexisting liquid and gas phases \cite{SammullerNeuralDensityFunctional2025}.
In particular, training on the basis of Eq.~\eqref{eq:EL_loss} without chemical potential values causes no noticeable decrease in the accuracy of the neural functional as compared to direct training via Eq.~\eqref{eq:c1_EL} with known values of $\mu_k$.
We recall that both methods deliver neural functionals that remain robust for predicting fluid equilibria not featured explicitly in the training dataset, such as the shown case of liquid-gas coexistence \cite{SammullerNeuralDensityFunctional2025}.
Further results for the hard sphere and isothermal Lennard-Jones fluids are presented in Appendix~B.
Code and data to reproduce the findings of this work are openly accessible~\cite{SammullerNeuralMu2025}.

\begin{figure}[tb]
  \includegraphics{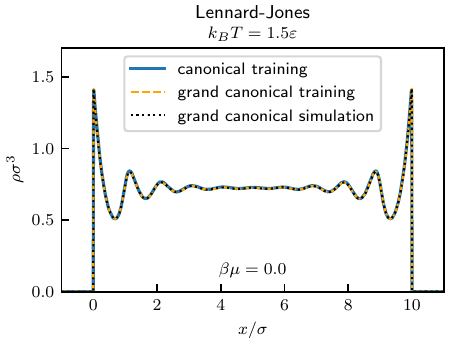}
  \caption{
    Results for the density profile (solid blue line) of the truncated Lennard-Jones fluid in planar hard wall confinement obtained from training a neural functional with data from canonical Monte Carlo simulations with unknown chemical potential values.
    For comparison, we have also performed training with grand canonical data (dashed orange line, see also Appendix~B and Fig.~\ref{fig:3D-hard-spheres_3D-LJ-T1.5}) and take as ultimate reference the density profile from a grand canonical Monte Carlo simulation (dotted black line).
    The agreement of all three routes confirms the accuracy of the prediction and the feasibility of neural density functional construction based on canonical data.
  }
  \label{fig:canonical}
\end{figure}

We next turn to canonical training and consider the truncated Lennard-Jones fluid at fixed temperature $k_B T = 1.5 \varepsilon$ \cite{SammullerNeuralFunctionalTheory2023,SammullerNeuralDensityFunctionals2024}.
The dataset comprises $n = 400$ individual canonical Monte Carlo simulation runs with randomized external potential landscapes and randomized particle numbers in the range $100 < N < 1500$ yielding mean densities $0.05 < \bar{\rho} \sigma^3 < 0.75$.
Due to the sufficiently large system size $10 \times 10 \times 20 \sigma^3$ and large particle numbers, we expect ensemble differences in the density profiles to be negligible and hence to obtain a suitable dataset for determining a universal functional mapping $c_1(x; [\rho])$.
The training proceeds on the available data for $V_\mathrm{ext}^{(k)}(x)$ and the canonical density profiles $\rho_k(x)$, where $k = 1, \dots, n$, and we observe that the minimization of the loss function~\eqref{eq:EL_loss} converges.
As no reference for chemical potential values is available, we perform an indirect verification by testing the predictions of the neural density functional $c_1(x; [\rho])$, reasoning that if this functional mapping is represented correctly, then the chemical potential values must follow accordingly from the successful minimization of Eq.~\eqref{eq:EL_loss}.
Figure~\ref{fig:canonical} shows a self-consistent density profile for planar hard-wall confinement obtained via the canonically trained neural functional.
For comparison, we also show the prediction of a neural functional trained on grand canonical data (see Appendix~B) as well as a density profile from direct grand canonical Monte Carlo simulation.
All results agree with each other, thus confirming the feasibility of constructing neural density functionals based on canonical simulation data.

In summary, the presented method constitutes an efficient means of determining chemical potential values for a given dataset of $n$ inhomogeneous fluid states.
The sole input quantities are the temperatures $T_k$, the external potentials $V_\mathrm{ext}^{(k)}(\vec{r})$ and the corresponding density profiles $\rho_k(\vec{r})$ of each system $k = 1, \dots, n$, which are directly accessible already in canonical simulation methods such as molecular or Brownian dynamics.
The viability of our framework fundamentally rests on the existence and uniqueness of the classical density functional mapping $T, \rho(\vec{r}) \rightarrow V_\mathrm{ext}(\vec{r})$ \cite{EvansNatureLiquidvapourInterface1979}.
In principle, a dataset of these quantities for various inhomogeneities should therefore be sufficient for recovering the many-body physics of the fluid model under consideration, implying that unknown chemical potentials can formally be determined a posteriori.
To put this task into practice, we have exploited the universal nature of the one-body direct correlation functional $c_1(\vec{r}; [\rho], T)$.
Density functional learning of a neural-network-based representation of $c_1(\vec{r}; [\rho], T)$ via the loss function \eqref{eq:EL_loss} then yields the missing system-specific chemical potentials $\mu_k$ essentially as a mere additive offset.
We reiterate that not a single value of $\mu_k$ needs to be known a priori, and that our first-principles-based method is conceptually different from more common applications of machine learning, where the underlying dataset usually features the target quantity directly.

The application of our method is particularly beneficial for large datasets, which arise for instance in solvation studies that are relevant for various biochemical processes and for computer-aided drug discovery \cite{MobleyFreeSolvDatabaseExperimental2014,LuukkonenPredictingHydrationFree2020,SergiievskyiFastComputationSolvation2014,RatkovaSolvationThermodynamicsOrganic2015}.
Direct means of measuring the chemical potential, e.g.\ using Widom's insertion method, incurs additional computational cost that scales linearly with the number $n$ of considered systems, which might be prohibitive for such data-intensive tasks.
In contrast, our method rests only on the availability of density profiles, from which the chemical potentials as well as further results can be obtained without resorting back to many-body simulation techniques.
As merely a machine-learning-based post-processing step is required, there is no relevant computational overhead besides the comparatively inexpensive density functional learning via Eq.~\eqref{eq:EL_loss}.
Recall further that the trained neural functional can readily be utilized for predictions of various thermodynamic and structural properties \cite{SammullerNeuralFunctionalTheory2023,SammullerNeuralDensityFunctional2025}, such as (solvation) free energies.

We conclude with several practical aspects and conceptual points.
Due to the underlying mechanism of density functional learning, the functional mapping needs to be probed sufficiently by the provided dataset such that training of a neural functional succeeds.
Datasets with only very few systems or with weak inhomogeneities may hamper this machine learning task; the reliability of predictions for reduced dataset sizes is investigated in Appendix~C.
Pair-correlation regularization \cite{DijkmanLearningNeuralFreeEnergy2025,SammullerNeuralDensityFunctionals2024} could remedy the limitations of insufficient inhomogeneous data to a certain extent by additionally matching the neural functional to two-body correlation functions in the bulk fluid.
Lastly, we emphasize that the temperature is required as a known quantity in the dataset.
It would be interesting in future work to evaluate the feasibility of using purely microcanonical data for training neural functionals and for recovering temperature values \cite{BinderMolecularDynamicsSimulations2004}, but also to investigate ensemble differences and canonical density functionals \cite{DeLasHerasFullCanonicalInformation2014} as well as connections to metadensity functional theory \cite{KampaMetadensityFunctionalTheory2025}.

\begin{acknowledgments}
  \emph{Acknowledgments}---%
  We thank Robert Evans for valuable discussions.
  Some of the calculations were performed using the emil- and festus-clusters of the Bayreuth Center for High Performance Computing funded by the DFG (Deutsche Forschungsgemeinschaft) under Project Nos.\ 422127126 and 523317330.
  This work is supported by the DFG (Deutsche Forschungsgemeinschaft) under Project No.\ 551294732.
\end{acknowledgments}

\emph{Data availability}---%
The data that support the findings of this Letter are openly available \cite{SammullerNeuralMu2025}.

\bibliography{bibliography}

\section{End matter}

\begin{figure}[tb]
  \includegraphics{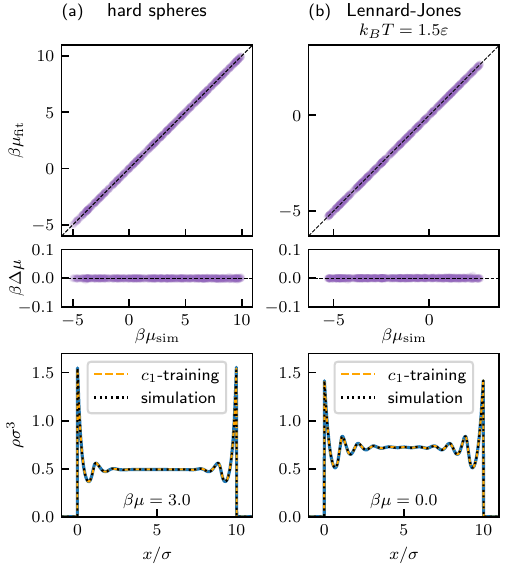}
  \caption{
    Results for the hard sphere fluid (a) and the truncated Lennard-Jones fluid at temperature $k_B T = 1.5 \varepsilon$ (b) analogously to Fig.~\ref{fig:1D-hard-rods_3D-LJ-Tvar}.
    The values $\beta \mu_\mathrm{fit}$ recovered via the present procedure agree with the simulation reference $\beta \mu_\mathrm{sim}$ (top row), as confirmed by the small values of the scaled potential difference $\beta \Delta \mu = \beta \mu_\mathrm{fit} - \beta \mu_\mathrm{sim}$ (middle row).
    The density profiles (bottom row, solid blue lines) correspond to confinement between hard walls of separation distance $10 \sigma$ and are obtained via self-consistent solution of Eq.~\eqref{eq:EL} with the neural functional $c_1(x; [\rho])$ resulting from the minimization of the loss function~\eqref{eq:EL_loss}.
    For comparison, we show results of neural functionals that have been trained directly with grand canonical data using known values of $\mu$ (dashed orange lines) as well as density profiles from grand canonical Monte Carlo simulations (dotted black lines).
  }
  \label{fig:3D-hard-spheres_3D-LJ-T1.5}
\end{figure}

\begin{figure}[tb]
  \includegraphics{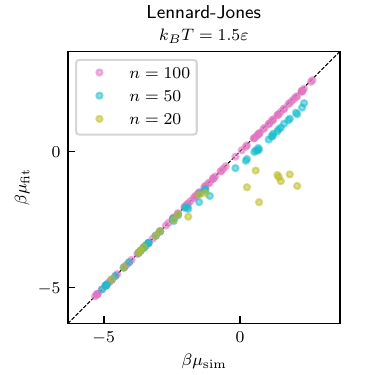}
  \caption{
    Comparison of results for training with reduced dataset sizes $n = 100$, 50, and 20, as indicated.
    Scaled chemical potential values $\beta \mu_\mathrm{fit}$ are recovered reliably for the case $n = 100$.
    Deviations to the reference data $\beta \mu_\mathrm{sim}$ become noticeable for $n = 50$.
    For $n = 20$, the procedure only yields correct results for small chemical potentials, while larger values are underestimated.
  }
  \label{fig:reduce_size}
\end{figure}

\emph{Appendix A: Density functional learning}---%
In the following, we give an overview of machine learned density functionals, focusing in particular on a local learning scheme for the one-body direct correlation functional \cite{SammullerNeuralFunctionalTheory2023,SammullerNeuralFunctionalTheory2023a,SammullerWhyNeuralFunctionals2024,SammullerNeuralDensityFunctionals2024,SammullerNeuralDensityFunctional2025}.
The key element of density functional learning is a neural network that represents a specific density functional relationship.
The one-body direct correlation functional $c_1(\vec{r}; [\rho], T) = -\delta \beta F_\mathrm{exc}([\rho], T) / \delta \rho(\vec{r})$ thereby arises as the arguably most natural choice, and it is defined formally as the (scaled) functional derivative of the excess free energy functional $F_\mathrm{exc}([\rho], T)$.
The neural functional theory by \citeauthor{SammullerNeuralFunctionalTheory2023} \cite{SammullerNeuralFunctionalTheory2023,SammullerNeuralFunctionalTheory2023a,SammullerWhyNeuralFunctionals2024,SammullerNeuralDensityFunctional2025} rests on representing $c_1(\vec{r}; [\rho], T)$ locally by a standard multilayer perceptron, i.e.\ a fully-connected feed-forward neural network with multiple hidden layers.
Alternatively, it has been proven feasible to represent $F_\mathrm{exc}([\rho], T)$ via various neural network architectures \cite{SammullerNeuralDensityFunctionals2024,DijkmanLearningNeuralFreeEnergy2025} and to obtain $c_1(\vec{r}; [\rho], T)$ in applications from automatic differentiation.
Neural density functionals have been shown to facilitate robust and systematic predictions of thermodynamic and structural properties for both homogeneous and heterogeneous fluids in multiscale settings and for states that have not been featured explicitly in the training data \cite{SammullerNeuralFunctionalTheory2023,SammullerWhyNeuralFunctionals2024,SammullerNeuralDensityFunctional2025}.

Choosing $c_1(\vec{r}; [\rho], T)$ as the target quantity is motivated by the physical insight that direct correlation functions remain short-ranged for short-ranged interparticle interactions \cite{HansenTheorySimpleLiquids2013}, such that a neural network may be constructed to represent the density functional relationship locally:
For given position $\vec{r}$, only density information from positions $\vec{r}'$ within a narrow window $|\vec{r}-\vec{r}'| < r_w$ is used to generate the value $c_1(\vec{r}; [\rho], T)$.
The neural network hence mirrors the intended input-output mapping
\begin{equation}
  \left.\rho(\vec{r}')\right|_{|\vec{r} - \vec{r}'| < r_w}, T \rightarrow c_1(\vec{r}).
\end{equation}
It is typically sufficient to set the spatial cutoff $r_w$ to values of the order of the particle size $\sigma$, such as e.g.\ $r_w = 2.56\sigma$ \cite{SammullerNeuralFunctionalTheory2023} for hard spheres of diameter $\sigma$ and $r_w = 3.5\sigma$ \cite{SammullerNeuralDensityFunctionals2024,SammullerNeuralDensityFunctional2025} for the truncated Lennard-Jones fluid.
Local learning has multiple benefits \cite{SammullerNeuralFunctionalTheory2023,SammullerNeuralFunctionalTheory2023a,SammullerWhyNeuralFunctionals2024,SammullerNeuralDensityFunctionals2024,SammullerNeuralDensityFunctional2025}, including the decoupling of predictions from the original simulation box size, which facilitates the subsequent utilization of the trained neural functional in multiscale applications.
We note that reformulations of the local learning scheme in terms of convolutional neural networks are possible, and that such approaches have recently been applied to the two-dimensional hard disk fluid \cite{GlitschNeuralDensityFunctional2025}.

\emph{Appendix B: Hard spheres and isothermal Lennard-Jones fluid}---%
We consider in the following the hard sphere fluid with particle diameter $\sigma$ and the truncated ($r_c = 2.5 \sigma$) Lennard-Jones fluid at constant temperature $k_B T = 1.5 \varepsilon$, and attempt to recover chemical potential values across corresponding datasets via the presented method.
As reference, we use grand canonical Monte Carlo simulations with planar inhomogeneous external potentials $V_\mathrm{ext}^{(k)}(x)$ and known chemical potentials $\mu_k$.
Analogously to the main text, the latter are however not prescribed and instead determined during training.
The total number of simulations is $n = 450$ for the athermal hard sphere fluid \cite{SammullerNeuralFunctionalTheory2023} and $n = 500$ for the isothermal Lennard-Jones system \cite{SammullerNeuralDensityFunctionals2024}.
For both fluid types, minimization of the loss function \eqref{eq:EL_loss} recovers accurately the chemical potential values of the entire datasets.
Notably each resulting neural functional $c_1(x; [\rho])$ delivers the same level of accuracy as the corresponding one that has been trained with prescribed values of $\mu_k$ according to Eq.~\eqref{eq:c1_EL}, which we recall requires full grand canonical information \cite{SammullerNeuralFunctionalTheory2023,SammullerNeuralDensityFunctionals2024}.
The chemical potential values and neural network predictions are assessed in Fig.~\ref{fig:3D-hard-spheres_3D-LJ-T1.5}.
See Ref.~\cite{SammullerNeuralFunctionalTheory2023} for a comparison of neural functional results to analytical treatments \cite{HansenGoosDensityFunctionalTheory2006}.

\emph{Appendix C: Reduced dataset}---%
We investigate the accuracy of the chemical potential predictions when the dataset size is decreased.
Figure~\ref{fig:reduce_size} shows a comparison for the isothermal Lennard-Jones system, as analyzed for the full dataset of $n = 500$ simulations in Appendix~B and Fig.~\ref{fig:3D-hard-spheres_3D-LJ-T1.5}(b).
Reduced dataset sizes of $n = 100$, 50, and 20 are considered by including only a subset of all simulation results in the training.
For $n = 100$ individual simulations, all their chemical potential values can still be recovered reliably.
Minor deviations to the reference values start to appear particularly for larger chemical potentials for $n = 50$.
If only $n = 20$ simulations are considered, recovering chemical potentials remains possible for small values of $\mu$, but the quality of the results worsens for larger chemical potential values, which tend to be underestimated.
This behavior is rationalized by recalling the underlying mechanism of the procedure, which hinges on the successful determination of the functional mapping $c_1(\vec{r}; [\rho], T)$.
For large values of $\mu$, the fluid is dense and highly correlated, such that it becomes particularly intricate to capture accurately the density functional dependence in this regime with only few training samples.

\end{document}